# Quantifying and attenuating pathologic tremor in virtual reality

Brian A. Cohn[*,1], Dilan D. Shah[*,2], Ali Marjaninejad[*,3], Martin Shapiro[*,4], Serhan Ulkumen[*,5,8],
Christopher M. Laine[*,6], Francisco J. Valero-Cuevas[3,6], Kenneth H. Hayashida[4], Sarah Ingersoll[7,‡]

*Abstract*— We present a virtual reality (VR) experience that creates a research-grade benchmark in assessing patients with active upper-limb tremor, while simultaneously offering the opportunity for patients to engage with VR experiences without their pathologic tremor. Accurate and precise use of handheld motion controllers in VR gaming applications may be limited for patients with upper limb tremor. In parallel, objective tools measuring tremor are not in widespread, routine clinical use. We used a commercially available VR system and designed a challenging virtual-balloon-popping test mimicking a common nose-to-target pointing task used by medical practitioners to subjectively evaluate tremor in the exam room. Within our VR experience, we offer a software mode which uses a low-pass filter to adjust hand position and pointing orientation over a series of past data points. This digital filter creates a smoothing function for hand movement which effectively removes the patient's tremor in the VR representation. While the patient completes trials of the reaching task, quantitative data on the pathologic tremor is digitally recorded. With speed, accuracy, and the tremor components computed across three axes of movement, patients can be evaluated for their tremor amplitudes in a quantitative, replicable, and enjoyable manner. Removal of tremor in digital space may allow patients having significant upper limb tremor to have both an objective clinical measurement of symptoms while providing patients positive feedback and interaction.

## I. Introduction

For patients living with Parkinson's Disease (PD) and Essential Tremor (ET) manipulation of objects and controls challenges their ability (and disability) with impact on identity and agency. We explore two areas of interest to this patient group with respect to Virtual Reality (VR) experiences. First, we examine whether a common reaching task used by clinicians (the finger-nose-finger test) could be automated in VR. Second, we explore whether software could be used to visually dampen the tremor while performing the VR task.

Rating scales (such as the UPDRSIII and the Fahn-Tolosa-Marin) and a variety of EMG-based measurement approaches have advanced the precision with which researchers can classify PD and ET patients [1]. Prior work has explored the 3D tracking of patient hand position over the course of functional movements [2]. However, these methods require custom-designed systems which implement electromyographic (EMG) sensors with 3D position tracking and have a level of complexity unconducive to widespread implementation [3]. Alternatively, a standardized VR tool could incorporate the experimental paradigm and proctorship within the experience itself. This may contribute to stronger replicability and consistency across subjects [4].

Existing VR systems are engineered for low-latency and high-precision tracking. Therefore, VR systems on the consumer market allow for immersive experiences. This research seeks to evaluate if the high-quality, high-accessibility tracking systems present in consumer VR devices could be applicable in the clinical, medical and rehabilitative management of patients with tremor. These VR systems and software could be rapidly deployed to enhance the quality of life of patients with neurological diagnoses, including PD and ET [5], [6].

## II. Methods

We produced a VR environment (Unity 2018, San Francisco, CA, USA) for use with a tracked head-mounted display (HMD) and one hand controller (HMD software-accessible rate of 90Hz; HTC Vive, Xindian, New Taipei City, Taiwan). Hand position and orientation for one controller was saved to RAM and saved to a comma-separated-value (CSV) file after the individual user trial was complete. The second hand was tested in a separate trial with the same controller.

### A. Task

The subject is instructed to keep their torso stationery against a high-backed chair while wearing the HMD and holding the controller in one of their hands. In VR, the subject sees a digital hand holding a pointy pencil in a position correlating to their own hand in space. An alternating, sequential presentation of two balloons is presented. One balloon appears in the same relative position in front of their nose. A second balloon appears as a reach target spawned in a random position along a line at a comfortable reaching distance for the subject (Figure 1). The patient moves between the two targets, popping each balloon with the pencil tip that is affixed to their VR-hand avatar. The length and position of the 'reach balloon' line can be adjusted and saved, allowing for a subject-specific and replicable test environment. Axes are illustrated in Figure 2.

A 'debounce' of 100ms was added to the balloon targets; in effect, we did not register a pop of the balloon unless the pencil tip was within the balloon collider for 100ms frames. This was added to prevent the user from simply swiping over or slashing into the targets, thereby promoting accuracy and stability at

[*]Authors contributed equally to work presented herein.

University of Southern California Departments of Computer Science[1], Biomedical Engineering[3], Keck School of Medicine[4], Department of Biological Sciences[5], Division of Biokinesiology and Physical Therapy[6], and Department of Neurology[7]

[2] Projectfutur.es
[8] Digital Dreams

[‡]Corresponding Author; singerso@usc.edu

each of the targets and mimicking the default behavior in the existing clinical test.

*B. Tremor elimination in VR*

A keyboard toggle can enable or disable smoothing of the controller position in the virtual environment. This is achieved by caching the past *K* controller states across the past *K* frames and defining the current virtual hand position as the mean position and orientation of the cache. Although large values of *K* cut all signs of tremor even below physiologically observed tremor frequencies, more filtering leads to controller lag. Ad-hoc calibration with the user allows for a comfortable level of filtering that allows for maximum responsiveness, given the elimination of tremor.

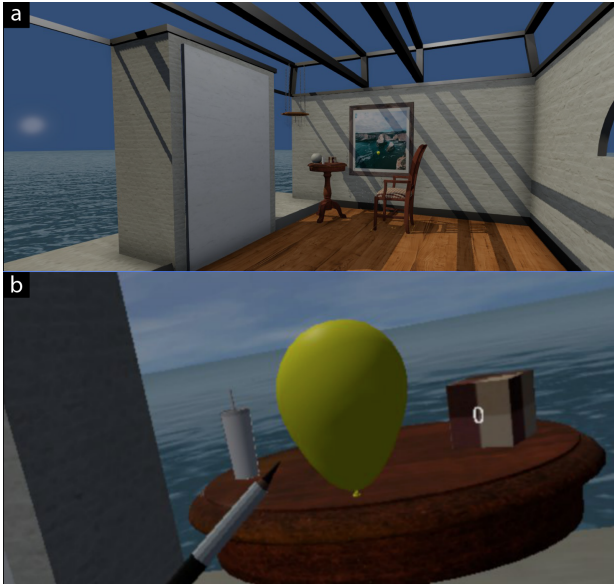

Figure 1. In-game view of the environment balloon task. In reality, the seated subject reaches repeatedly from their nose to a target suspended ahead of them, and 3D tracking is achieved through inside-out HMD and controller tracking with two infrared-light-emitting base stations. (a) The room is a dimly-lit room with a virtual chair that is lined up with a stationary physical chair in reality. (b) The balloon is popped by holding the pencil tip in the balloon's volume. The score thus far is visible to the user.

*C. Analysis*

Python 3.6 was used to produce visualizations of the recorded data and compute frequency distributions for each of the X, Y, and Z rotational axes for the controller. With the Python script running within the environment, trial-specific data (much like data shown in this paper) are immediately visible on the wall of the virtual environment and are concatenated into a PDF for a clinical provider's review with the patient. Discrete Fourier Transform (DFT) was applied on data on each axis to provide frequency domain analysis of the tremor. We do not perform any further analyses, though we acknowledge that extended techniques are highly applicable (i.e. as with asymmetry analysis, and interpretation of the endpoint positions, velocities, and accelerations) [7].

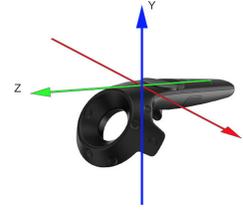

Figure 2. Orientation of the three axes of interest for this study. The pencil tip was placed along the Z dimension, central to the controller. Image from msi.com/blog/getting-started-with-mixed-reality

## III. RESULTS

Tremor is clearly observable in the frequency domain when investigating each of the three rotational angles over the course of 15 reaching trials (15 reach balloons, and 15 nose balloons in total). All data shown is collected from one subject across two conditions of non-tremorous and tremorous action. For the tremorous condition, the subject was instructed to produce a 4-7Hz tremor, with a metronome as a guide for guiding their timing. With a Hanning window of 270 samples over the data collected at 90Hz, and with an overlap of 135 samples, we computed a Discrete Fourier Transform over frequencies from 2-12Hz, a generous range for human-observed tremor frequency bands [1]. Performing this analysis for each of the controller rotations in *x*, *y*, and *z*, and across the two conditions of non-tremorous and tremorous movement, frequency components in the 4-6Hz band are clearly visible (Figure 3).

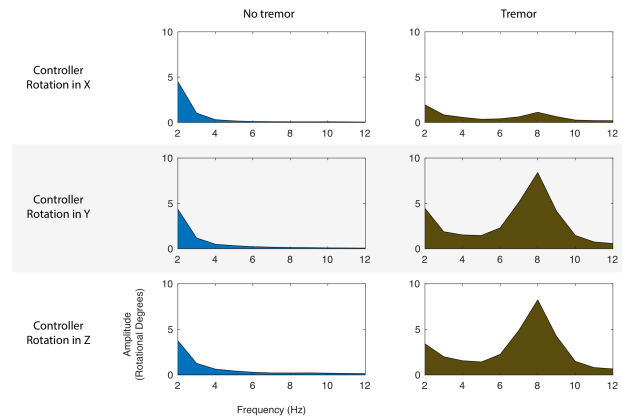

Figure 3. Data collected for a subject creating a patient-like tremorous movement. (a) 3D positions over time are used to generate (b) frequency distribution curves, which illustrate the amplitude of shaking across the different frequencies, along different axes of movement.

## IV. CONCLUSION

We find preliminary evidence that primary upper limb movements of PD and ET can be quantified through a VR reaching paradigm. Moreover, functionally disruptive tremulous frequencies can be attenuated via a visual moving-average filter. This work provides technical precedent and a scientific proof-of-principle for clinical applications.

ACKNOWLEDGMENT

We thank HTC for their sponsorship award and provision of a VR system for use with this research, the National Science Foundation Graduate Research Fellowship (NSF GRF) to B.C., Provost and Research Enhancement Fellowships from the graduate school of University of Southern California (USC) to A.M, NIH Grants R01-052345 and R01-050520, and by the Department of Defense under award number MR150091 to F.V-C.